\documentstyle[aps,epsfig]{revtex}
\begin{document}
\draft
\tighten
\title{Transverse  energy distributions and $J/\psi$ production
in Pb+Pb collisions}
\author{\bf A. K. Chaudhuri\cite{byline}}
\address{ Variable Energy Cyclotron Centre\\
1/AF,Bidhan Nagar, Calcutta - 700 064\\}
\date{\today}

\maketitle

\begin{abstract}
We  have  analyzed  the  latest  NA50  data  on transverse energy
distributions and $J/\psi$ suppression in Pb+Pb  collisions.  The
transverse  energy  distribution  was  analysed  in the geometric
model of AA collisions. In the geometric model,  fluctuations  in
the  number  of NN collisions at fixed impact parameter are taken
into  account.  Analysis  suggests  that  in  Pb+Pb   collisions,
individual  NN collisions produces less $<E_T>$, than in other AA
collisions.  The  nucleons  are   more   transparent   in   Pb+Pb
collisions.  The  transverse  energy  dependence  of the $J/\psi$
suppression was obtained following the model of  Blaizot  et  al,
where  charmonium  suppression  is  assumed to be 100\% effective
above a threshold density. With  fluctuations  in  number  of  NN
collisions  taken into account, good fit to the data is obtained,
with a single parameter, the threshold density.
\end{abstract}


In  relativistic  heavy ion collisions $J/\psi$  suppression has
been recognized as an important tool  to  identify  the  possible
phase  transition  to   quark-gluon plasma. Because of the large
mass of the charm quarks, $c\bar{c}$  pairs  are  produced  on  a
short  time  scale.  Their tight binding also make them immune to
final state interactions. Their evolution  probes  the  state  of
matter in the early stage of the collisions.   Matsui
and  Satz  \cite{ma86}  predicted that in presence of quark-gluon
plasma (QGP), binding of $c\bar{c}$  pairs  into  $J/\psi$  meson
will  be  hindered, leading to the so called $J/\psi$ suppression
in heavy ion  collisions  \cite{ma86}.  Over  the  years  several
groups  have measured the $J/\psi$ yield in heavy ion collisions
(for a review of the data and the interpretations see
\cite{vo99,ge99}).   In   brief,   experimental   data   do   show
suppression.  However this  could   be   attributed   to   the
conventional  nuclear  absorption,  also  present  in pA
collisions.

The latest data obtained by the NA50 collaboration \cite{na50} on
J/$\psi$ production in Pb+Pb collisions at 158 A GeV is the first
indication  of  anomalous  mechanism  of  charmonium suppression,
which  goes  beyond  the  conventional  suppression  in   nuclear
environment.   The   ratio   of   J/$\psi$   yield to that of Drell-Yan pairs
 decreases faster with $E_T$ in  the  most  central
collisions  than  in the less central ones. It has been suggested
that the resulting pattern can be understood in  a  deconfinement
scenario  in terms of  successive melting of charmonium bound states \cite{na50}.

In  a recent paper Blaizot et al \cite{bl00} showed that the data
can be understood as an effect of transverse energy  fluctuations
in   central   heavy   ion   collisions.   Introducing  a  factor
$\varepsilon=E_T/E_T(b)$   and assuming  that
the  suppression  is 100\% above a threshold density (a parameter
in the  model) and
smearing  the  threshold  density  (at  the  expense  of  another
parameter) best fit to the  data  was  obtained.  Capella  et  al
\cite{ca00}   analysed the data in the comover approach. There
also,  the  comover  density  has  to  be  modified by the factor
$\varepsilon$. Introduction of this adhoc factor $\varepsilon$ can
be justified in a model based on excited nucleons represented  by
strings \cite{hu00}.

At  a  fixed  impact parameter,   the transverse energy as well as the
number of NN collisions     fluctuate. The  Fluctuations  in  the
number  of  NN  collisions  were  not  taken  into account in the
calculations of Blaizot et al \cite{bl00} or in the  calculations
of  Capella et al \cite{ca00}. In the present paper, we present a
calculation following the model  of  Blaizot  et  al  \cite{bl00} which
includes these  fluctuations.  As  will  be shown below, if the
fluctuations in number of NN collisions are  taken  into  account
very  good  description  of the NA50 data can be obtained without
smearing the threshold density. The smearing effect  is  generated  by
the fluctuations.

Geometric  model  has  been  quite  successful  in explaining the
transverse   energy   production   as   well   as    multiplicity
distributions  in  AA collisions \cite{ch90,ch93}. In this model,
it is assumed that at impact parameter ${\bf b}$, the number  $n$
of  NN  collisions  is  Poisson distributed with average $<n_b>$.
In the Glauber approximation $<n_b>$ is written as,,

\begin{equation}
<n_b>=\sigma_{NN}  \int  d^2s  T_A({\bf  s}) T_B({\bf s}-{\bf b})
\label{1}
\end{equation}

\noindent  where $\sigma_{NN}$  is  the  inelastic  NN  cross-section,
assumed to be 32 mb. In this model all the nuclear information is
contained in the  nuclear  thickness  function  is  $T_{A,B}({\bf
s})=\int  dz  \rho_{A,B}({\bf s},z)$. In the present calculation,
we have  used  the  following  parametric  form  for  $\rho_A(r)$
\cite{bl00},

\begin{equation}
\rho_A(r)=\frac{\rho_0}{1+exp(\frac{r-r_0}{a})}
\end{equation}

\noindent  with  a=0.53 fm, $r_0=1.1A^{1/3}$. The central density
is obtained from $\int \rho_A(r)d^3r=A$.

In the geometric model, the probability to obtain $E_T$ at impact
parameter  ${\bf  b}$  in $n$- number of NN collisions is written
as,

\begin{equation}
P_n(b,E_T)=     \frac{e^{-<n_b>}    <n_b>^n}    {\Gamma    (n+1)}
Q^{\{n\}}(E_T) \label{2}
\end{equation}

\noindent  where  $Q^{\{n\}}(E_T)$  is  the n-fold convolution of
$E_T$ distribution resulting from elementary NN collisions,

\begin{equation}\label{3}
Q^{\{n\}}(E_T)=   \int   dE^1_T   ...   dE^n_T
g(E^1_T)...g(E^n_T) \times  \delta(E_T-E^1_T...-E^n_T)
\end{equation}

In  eq.\ref{3}  $g(E_T)$ is the normalized $E_T$ distribution for
NN collisions. Most of the $E_T$ distributions in  NN  collisions
can be well approximated by the gamma distribution,

\begin{equation}  g(E_T)=  \frac{\alpha^\beta}{\Gamma  (\beta  )}
e^{-\alpha E_T} E^{\beta -1}_T \end{equation}

\noindent with the parameters $\alpha$ and $\beta$. For the gamma
distribution, the average and the variance are,

\begin{mathletters}
\begin{eqnarray}
<E_T>_{NN}= && \beta/\alpha          \\
<E^2_T>/<E_T>^2-1=&&  1/\beta
\end{eqnarray}
\end{mathletters}

Gamma  distribution  has  an  elegant  convolution property which
greatly facilitate computation. n-fold  convolution  of  a  gamma
distribution  is  again  a  gamma  distribution,  with parameters
$\alpha^\prime=\alpha$ and $\beta^\prime=n \beta$. Thus,

\begin{equation}
Q^{\{n\}}(E_T) = \frac{\alpha^{n \beta}}{\Gamma  (n \beta  )}
e^{-\alpha E_T} E^{n \beta -1}_T
 \end{equation}

The  final  transverse  energy distribution is then obtained from
eq.\ref{2} by summing it  over  n  (from  1  to  $  \infty  $)  and
averaging over the impact parameter {\bf b}.

Geometric model has been quite successful in explaining the $E_T$
distributions  in  heavy  ion collisions with $\alpha \sim$ 2 and
$\beta \sim$ 2 \cite{ch93}. We have fitted the $E_T$ distribution
in Pb+Pb collisions varying the parameter $\alpha$  and  $\beta$.
NA50  collaboration  did  not  correct  the $E_T$ spectra for the
efficiency of target identification  algorithm,  which  is  lower
than  unity for $E_T$ lower than 60 GeV. To obtain the parameters
$\alpha$ and $\beta$ we have fitted the purely inclusive part  of
the  $E_T$ spectra ($E_T >$ 60 GeV). Very good fit to the data is
obtained with $\alpha= 3.46 \pm .19$ and $\beta=0.379 \pm 0.021$.
The fit is shown in fig.1. We note that best fitted values of
$\alpha$ and $\beta$ indicate  that  average  $E_T$  produced  in
individual  NN  collisions  is order of magnitude smaller than in
other AA collisions \cite{ch93}. The variance is also a order  of
magnitude  large. It seems that the nucleons are more transparent in
Pb+Pb collisions than in (say) O+Au collisions, leading  to  less
$<E_T>$  in  individual  NN  collisions. It is evident that $E_T$
production mechanism in Pb+Pb collisions is different from  other
AA  collisions.  Transparency of nucleons in Pb+Pb collisions may
be interpreted as  an  indication  of  QGP  production  in  Pb+Pb
collisions.

\begin{figure}[h]
\centerline{\psfig{figure=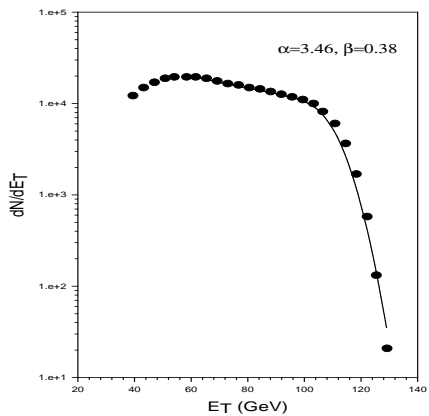,height=6cm,width=6cm}}
\caption{Transverse  energy  distribution  in  Pb+Pb  collisions}
\end{figure}

As  mentioned  in the beginning, we have followed the model of Blaizot
et al \cite{bl00} to analyze the $E_T$ dependence  of  charmonium
suppression.   Charmonium   production  cross-section  at  impact
parameter ${\bf b}$ is written as,

\begin{equation}
d^2\sigma^{J/\psi}/d^2b     =     \sigma^{J/\psi}    \int    d^2s
T^{eff}_A({\bf s}) T^{eff}_B({\bf s}-{\bf b}) S({\bf b},{\bf  s})
\label{6b}
\end{equation}

\noindent   where   $T_{A,B}^{eff}$   is  the  effective  nuclear
thickness function,

\begin{equation} \label{6c}
T^{eff}      ({\bf s})=\int_{-\infty}^\infty dz \rho({\bf s},z)
exp(-\sigma_{abs}\int_z^\infty dz^\prime \rho({\bf s},z^\prime))
\end{equation}

\noindent with  $\sigma_{abs}$  as  the  cross-section for $J/\psi$
absorption by nucleons. The exponential  factor  is  the  nuclear
absorption   survival   probability,   the  probability  for  the
$c\bar{c}$ pair to avoid nuclear absorption and form a  $J/\psi$.
$S({\bf  b},{\bf  s})$  is the anomalous part of the suppression.
Blaizot et al \cite{bl00} assumed that  $J/\psi$  suppression  is
100\% effective above a threshold density ($n_c$), a parameter in
the model. Accordingly the anomalous suppression part was written
as,

\begin{equation}
S({\bf b},{\bf  s})= \Theta(n_c - n_p({\bf b},{\bf s})) \label{7}
\end{equation}

\noindent  where  $n_p$ is the density of participant nucleons in
impact parameter space,

\begin{equation}
n_p({\bf      b},{\bf     s})=     T_A({\bf
s})[1-e^{-\sigma_{NN}   T_B({\bf    b}-{\bf    s})}]    +    [T_A
\leftrightarrow T_B] \label{8} \end{equation}

Recognizing that the endpoint behavior of charmonium suppressions
are   due  to  transverse  energy  fluctuations,  Blaizot  et  al
\cite{bl00} modified the density of  participant  nucleons  by  a
factor  $\varepsilon=E_T/E_T(b)$.  This  modification  makes sense
only when $n_p$ is assumed  to  be  proportional  to  the  energy
density.   Implicitly   it   was  also  assumed  that  the  $E_T$
fluctuations are strongly correlated in different rapidity  gaps.
The assumption was essential as NA50 collaboration measured $E_T$
in  the  1.1-2.3  pseudorapidity window while the $J/\psi$'s were
measured in the rapidity window $2.82<y<3.92$ \cite{na50}. In the
geometric model, strong correlation between $E_T$ fluctuations in
different rapidity window is readily obtained. At a fixed  impact
parameter, fluctuations in $E_T$ can be calculated as \cite{ch93}

\begin{equation}
\frac{<E_T^2>_{AA}-<E_T>^2_{AA}}{<E_T>^2_{AA}}                  =
\frac{<N^2>-<N>^2}{<N>^2}             +             \frac{1}{<N>}
\frac{<E_T^2>_{NN}-<E_T>^2_{NN}}{<E_T>^2_{NN}}
\end{equation}

The  $E_T$ fluctuations has two parts, (i)  geometric in nature
which remains same irrespective of  rapidity  window, 
(ii) which  depend on fluctuations in the $E_T$ distributions in
NN collisions. The 2nd  part changes with  rapidity  window  but  its
effect  is  less  as  it  is  weighted  by  the  factor  $1/<N>$.
Fluctuations of $E_T$ in  different  rapidity  windows  are  thus
strongly correlated. 

We  calculate the $J/\psi$ production as a function of transverse
energy, at an impact parameter ${\bf b}$ as,

\begin{equation}\label{1a}
d \sigma  ^{J/\psi}/dE_T=\sum  _{n=1}^\infty
P_n(b,E_T) P(\psi \mid E_T,{\bf b})
\end{equation}

\noindent  where  $P_n(b,E_T)$ is the probability to obtain $E_T$
  in $n$ NN  collisions  (eq.(\ref{2}))
and  $P(\psi  \mid  E_T,{\bf b})$  is  the  probability  to  produce  a
charmonium with transverse energy $E_T$. 
 $P(\psi  \mid  E_T,{\bf b})$  is given by eq.\ref{6b}, with anomalous suppression  part
modified according to,

\begin{equation}
S({\bf  b},{\bf  s})  =  \Theta  (n_c - \frac{E_T}{n\beta/\alpha}
n_p({\bf b},{\bf s}))
\end{equation}

\noindent  where  we  have  replaced  $E_T(b)$ by $n\beta/\alpha$
appropriate in the geometric model. This modification takes  into
account  the  fluctuations  in  number  of NN collisions at fixed
impact parameter ${\bf b}$.

The  Drell-Yan  production  was  calculated  similarly, replacing
$P(\psi \mid E_T,{\bf b})$ in eq.\ref{1a} by the  Drell-Yan  production
cross-section,

\begin{equation}
d^2\sigma^{DY}/d^2b   =   \sigma^{DY}   \int   d^2s
T_A({\bf s}) T_B({\bf s}-{\bf b}) \label{6a}
\end{equation}

\begin{figure}[h]
\centerline{\psfig{figure=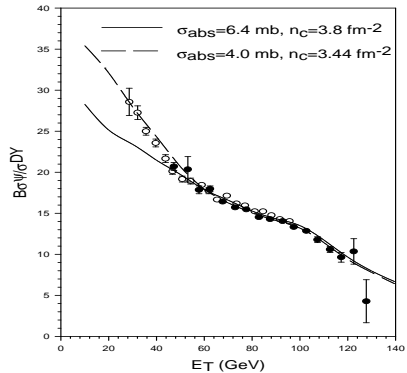,height=6cm,width=6cm}}
\caption{$J/\psi$  survival  probability  in Pb+Pb collisions as a
function of transverse energy.}
\end{figure}

In  fig.2, we have compared the theoretical charmonium production
cross-section with  NA50  experimental  data.  The  normalization
factor  $\sigma^{J/\psi}/\sigma^{DY}$  was  taken to be 53.5. The
solid curve is obtained with $\sigma_{abs}$=6.4 mb, and  $n_c=3.8
fm^{-2}$. Very good description of the data from 40 GeV onward is
obtained.  It  may  be  noted  that if the fluctuations in the NN
collisions were neglected,  equivalent  description  is  obtained
with  threshold  density  $n_c=3.75  fm^2$,  with smearing of the
$\Theta$ function at the expense  of  another  parameter.  It  is
evident  that  in this model, the smearing is done by fluctuating
NN collisions. Theoretical calculations predict more suppressions
below 40 GeV, a feature evident  in  other  models  also. 
It  is
possible  to fit the entire $E_T$ range, reducing the
$J/\psi$-nucleon absorption cross-scetion.  The  dashed  line  in
fig.2, corresponds to $\sigma_{abs}$=4 mb and $n_c=3.42 fm^{-2}$.

To summarize, we have analysed the transverse energy distribution
in  Pb+Pb  collisions  as  well as the charmonium production data
obtained  by  the  NA50  collaboration.  The  transverse   energy
distribution  was  analysed  in  the geometric model. It was seen
that in  order  to  fit  the  experimental  data,  individual  NN
collisions  are  required to produce (on the average) less $E_T$,
compared  to  other  AA  collisions.  It  seems  that  in   Pb+Pb
collision,   the  nucleons  become  transparent.  The  charmonium
production data was analysed following the model of Blaizot et al
\cite{bl00}, including the effect of fluctuations in number of NN
collisions at fixed impact parameter. The experimental data  from
40  GeV onwards could be very well fitted with $\sigma_{abs}$=6.4
mb and a threshold  density  of  3.8  $fm^{-2}$.  Neglecting  the
fluctuations  in  number  of  NN collisions, equivalent fit could
only be obtained by smearing the $\Theta$ function at the expense
of an added parameter. Data in the entire $E_T$  range  could  be
fitted  reducing  the  $J/\psi$-nucleon absorption cross-section to
$\sigma_{abs}$=4 mb. The threshold density  is  also  reduced  to
3.44 $fm^{-2}$. Considering that nucleons become more transparent
(as  suggested  by  the  $E_T$  data),  such  a  reduction  seems
plausible. Melting of charmoniums above a  threshold  density  as
well  as  apparent  transparency  of  nucleons  (as  evident from
analysis of $E_T$ distribution) strongly suggests that  in  Pb+Pb
collisions, QGP like environment is produced.

\end{document}